\documentclass[12pt,preprint]{aastex}

\shortauthors{Valle et al.}
\shorttitle{Evolution of Li, Be and B}

\newcommand{\beq}{\begin{equation}}
\newcommand{\enq}{\end{equation}}

\received{2001 July 17}
\begin{document}

\title{Evolution of Li, Be and B in the Galaxy}

\author{Giada Valle and Federico Ferrini}
\affil{Dipartimento di Fisica \\
Universit\`a di Pisa \\
Piazza Torricelli 2 \\
I-56100 Pisa, Italy}
\email{giada,federico@astr2pi.difi.unipi.it}

\author{Daniele Galli}
\affil{Osservatorio Astrofisico di Arcetri \\
Largo Enrico Fermi 5 \\
I-50125 Firenze, Italy}
\email{galli@arcetri.astro.it}

\author{Steven N. Shore}
\affil{Department of Physics and Astronomy \\
Indiana University South Bend \\
1700 Mishawaka Ave, South Bend \\
IN 46634-7111 USA\\
and\\
Osservatorio Astrofisico di Arcetri \\
Largo Enrico Fermi 5 \\
I-50125 Firenze, Italy}
\email{sshore@paladin.iusb.edu}

\begin{abstract}

In this paper we study the production of Li, Be and B nuclei by
Galactic cosmic ray spallation processes.  We include three kinds of
processes:  ({\em i}\/) spallation by light cosmic rays impinging on
interstellar CNO nuclei ({\em direct} processes); ({\em ii}\/)
spallation by CNO cosmic ray nuclei impinging on interstellar $p$ and
$^4$He ({\em inverse} processes); and ({\em iii}\/) $\alpha$--$\alpha$
{\em fusion} reactions.  The latter dominate the production of
$^{6,7}$Li.  We calculate production rates for a closed-box Galactic
model, verifying the quadratic dependence of the Be and B abundances
for low values of $Z$.  These are quite general results and are known
to disagree with observations.  We then show that the multi-zone
multi-population model we used previously for other aspects of Galactic
evolution produces quite good agreement with the {\em linear} trend
observed at low metallicities {\em without fine tuning}.  We argue that
reported discrepancies between theory and observations do not represent
a nucleosynthetic problem, but instead are the consequences of
inaccurate treatments of Galactic evolution.

\end{abstract}

%\keywords{Galaxy: evolution, --- cosmic rays --- spallation processes }

\section{Introduction}

The formation of the lightest elements beyond helium ($^{6,7}$Li,
$^9$Be, $^{10,11}$B, hereafter LiBeB) has persisted as a major problem
for nucleosynthesis for many decades.  The history is well known and we
will not review it here \citep[see][]{sppal,pts,ral99,fo99a,
fo99b,vfca,fovc00}. Recent observational advances have again brought
these elements to center stage as data become available for ancient,
extremely low metallicity stars in the Galactic halo and contemporary,
interstellar diffuse gas.

Several separate issues must be addressed regarding the time
development of light element abundances in the Galaxy.  The {\em Spite
plateau} appears to be the floor for $^7$Li and {\em may} represent its
primordial value.  Both isotopes of Li are easily destroyed in stellar
interiors so the likely source for the greater than primordial
abundances must be extra-stellar.  For this reason spallation reactions
by Galactic cosmic rays (GCR) interacting with nuclei in the diffuse
interstellar medium (ISM) have been implicated as avenues for light
element Galactic synthesis. Recent observations tax the simple
spallation models that have $p$ and $\alpha$ cosmic rays reacting with
stationary interstellar heavy nuclei.  As we will show, because the
heavy nuclei and massive cosmic rays come from the same ultimate source
-- stars and their supernova remnants -- the abundance of $^6$Li, Be,
and B should all scale as $Z^2$.  This prediction is severely at odds
with the observational data at low metallicity \citep{dptp98, bal99}.
This picture can be modified in several ways.  Inverting the reasoning,
\citet{p93} proposed that a change in the slope and lower energy cutoff
of GCRs over Galactic history can enhance the production of the light
nuclei and produce an almost linear dependence on $Z$. \citet{dal97}
argued that including fusion and inverse reactions, those for which the
high energy particles are accelerated CNO spalling stationary hydrogen
and helium in the ISM, will produce the right metallicity scaling for
even $^7$Li. \citet{fo99a} and \citet{fovc00} have shown that in a
closed box model the early evolution of LiBeB can be explained using
the standard rates provided the O/Fe ratio is allowed to float with
time, increasing at low $Z$.  These calculations were all performed
assuming only single galactic zones, albeit including simple infall
prescriptions.

Our aim in this paper is to contrast the predictions of two broad
classes of schemes for treating Galactic chemical and population
evolution:  simple closed box models, and more developed multi-zone,
multi-population models.  We will show that many of the theoretical
puzzles noted in the literature are simply paradoxes arising from
assumptions in the closed one-zone models.  These cannot be repaired by
more accurate prescriptions of the yields and delays due to stellar
evolution.  They are endemic to the basic feature of the models since
the system maintains constant mass without feedback and the star
formation is a monotonically decreasing function of time according to a
Schmidt-type power law in gas density (or surface density).

Once the nuclear reactions involved in the production of LiBeB nuclei
are established, a Galactic evolution model is needed to compute the
ISM abundances evolution and the SN rates. Two extreme choices are
possible:  the simple closed-box model \citep{t80}, or a more complex
nonlinear scheme, that gives satisfactory answers to other aspects of
Galactic evolution \citep{f91,fmpp92,par95}. A comparison between the
two is useful to fix the limits of influence of the hypotheses for the
GCR:  the observational situation of LiBeB reflects simply a nuclear
problem of GCR interactions with ISM, and hence depends exclusively on
knowledge about their interactions.

\section{Cosmic-ray nucleosynthesis}

In this section we consider the nucleosynthesis scenario of spallation
reactions of GCR on ISM nuclei for the production of LiBeB nuclei.
Among all nuclear reactions, there are two main production channels of
light elements:

({\em i}\/) {\em direct processes}, i.e. interactions of a $p$ or an $\alpha$
    particle of GCR with C,N,O nuclei of the ISM, hereafter represented
    by $p, \alpha$ + C,N,O $\rightarrow$ LiBeB;

({\em ii}\/) {\em inverse processes}, i.e. interactions of C,N,O nuclei of GCR
    with a $p$ or an $\alpha$ particle of the ISM, hereafter
    represented by C,N,O + $p, \alpha \rightarrow$ LiBeB.

For Li, {\em fusion reactions} must also be considered, i.e.
interactions of $\alpha$ particles of GCR with $\alpha$ particles of
ISM (hereafter represented by $\alpha + \alpha \rightarrow$
$^{6,7}$Li).

The production rate of the light element $\ell$ (with $\ell=$ LiBeB) at time
$t$, resulting from the spallation process $i+j\rightarrow \ell$, is the
convolution of the flux $\Phi_i(E_i,t)$ of the GCR projectile nuclei
$i$ with the numerical abundance of ISM target $x^{\rm ISM}_j$ weighted
by the spallation cross section $\sigma_{i,j\rightarrow \ell}(E_i)$,
\beq
\label{ratec}
r_{i,j\rightarrow \ell}(t)=x_{j}^{\rm ISM}(t)
\int_{E_T}^\infty\Phi_i(E_i,t)
\sigma_{i,j\rightarrow k}(E_i)S_{i,j\rightarrow \ell}(E_i)dE_i,
\enq
where $E_T$ is the threshold energy of the reaction.  

We assume that only a fraction $S_{i,j\rightarrow \ell}(E_i)$ of newly
produced light elements is thermalized and incorporated in the ISM, the
remaining fraction becoming part of the GCR. The reaction products Li,
Be and B have, in fact, different fates:  they can escape from the
interaction region if they have a sufficient energy, or suffer a
progressive energy loss until becoming thermalized.  The ability of the
ISM to stop the light elements so synthesized determines the survival
probability $S_{i,j\rightarrow \ell}(E_i)$ of various reaction products
against escape from the interaction region and for their nuclear
destruction that can occur during their slowing down. When the incident
particles are $p$ or $\alpha$ colliding with heavy nuclei,
$S_{i,j\rightarrow \ell}(E_i)\simeq 1$. In the opposite case, and for
$\alpha$--$\alpha$ reactions, the trapping fractions are usually
calculated in the framework of the {\em leaky-box} model \citep[see
{\em e.g.}][]{c80}, and have the form
\beq
S_{i,j\rightarrow \ell}(E_i)=\int f(E_\ell[E_i])
\exp\left[-\frac{R_\ell(E_\ell)}{\Lambda}\right]\;dE_\ell,
\enq
where $f(E_\ell[E_i])$ is the distribution of energy of the product
particle $\ell$ (depending on the energy $E_i$ of the incident particle),
$R_\ell(E_\ell)$ is the {\em stopping range} of the particle $\ell$, and
$\Lambda$ is the ISM path length.  Here we approximate the trapping
fractions as constant, and we adopt the results by \citet{m72}, shown
in Table \ref{tab:trapping}, that are in agreement with the
prescriptions by \citet{mar71} and other authors.

We assume that the GCR flux $\Phi_i(E,t)$ has a separable form,
\beq
\label{GCRspectrum}
\Phi_i(E,t)=x_i^{\rm GCR}(t)\xi(t)\varphi(E),
\enq
where $\varphi(E)$ is the present-day cosmic ray flux (number of nuclei
per unit energy time and area, see Sec. \ref{subsec:GCR}), $x_i^{\rm
GCR}(t)$ is the cosmic ray composition (by number), and $\xi(t)$ is a
dimensionless function that modulates the intensity of the GCR (see
Sec. \ref{subsec:GCR}), with $\xi=1$ at the present epoch.  Thus, eq.
(\ref{ratec}) becomes
\beq
\label{rijk}
r_{i,j\rightarrow \ell}(t)=x_i^{\rm GCR}(t)x_j^{\rm ISM}(t)
\xi(t){\cal P}_{i,j\rightarrow \ell} S_{i,j\rightarrow \ell},
\label{defrijk}
\enq
where we define the {\em production coefficients}
\beq
\label{pijk}
{\cal P}_{i,j\rightarrow \ell}=\int_{E_T}^\infty\varphi(E)
\sigma_{i,j\rightarrow \ell}(E)dE. 
\enq

The function $\xi(t)$ gives the time dependent intensity of GCR.
Following the current interpretation, we assume that cosmic rays
particles are accelerated by SN remnants, and therefore we set
\beq
\label{CRI1} 
\xi(t)=\frac{\psi_{\rm SN}(t)}{\psi_{\rm SN}(t_{\rm Gal})},
\enq
where $\psi_{\rm SN}$ is the global supernova rate (number of supernovae
per unit time) and $t_{\rm Gal}\simeq 13$~Gyr is the age of the Galaxy.

The composition of GCR varies with time according to the evolution of
interstellar abundances. Thus we set $x_{i}^{\rm GCR}(t)=x_i^{\rm
ISM}(t)$, where $x_i^{\rm ISM}(t)$ ($i=p$, $\alpha$, C, N, O) is the
abundance by number of the corresponding element in the ISM, calculated
by our model of Galactic chemical evolution.  Finally, the spallation
cross sections $\sigma_{i,j\rightarrow \ell}(E_i)$ for all direct 
and inverse processes are taken from \citet{rv84}, whereas the 
cross sections for $\alpha$--$\alpha$ reactions are from \citet{mal01}. 

\subsection{The energy spectrum of GCR}
\label{subsec:GCR}

A definitive, comprehensive theoretical framework is still lacking that
is able to explain the origin and mechanism(s) of propagation of
Galactic cosmic rays based on the interaction of relativistic charged
particles with the interstellar matter.  Approximate semiempirical
models are widely used to describe the propagation of GCR, often
treated as a diffusion process.  The ``leaky box'' model is the most
popular: a number of pointlike sources distributed throughout the
Galaxy emits a flux of fast particles with a broad energy spectrum. The
particles propagate diffusively within the Galaxy and have a certain
probability of escaping. The homogeneity hypothesis allows a simplified
treatment of the corresponding diffusion equation for the flux, and the
propagation equation can be solved assuming steady state conditions.
The solution depends on several factors:  the energy losses by
ionization, the mean free path for escape from the Galaxy, the mean
free path for nuclear reactions, the assumed spectrum at the source,
etc.  Energy loss resulting from continuous ionization accounts also
for the probability that a particle from the source or produced by
nuclear reactions during the propagation of GCR can survive to become
part of the flux.

In this work we adopt the propagated demodulated spectrum obtained by
\citet{ip85}, which we approximate as 
\beq
\varphi(E)=A(E+E_0)^{-\gamma},
\label{spectrum}
\enq
where $E$ is in MeV, $E_0=750$~MeV, and $\gamma=2.7$. For $E\gg E_0$
this spectrum becomes a simple power law in energy that allows to adopt
the absolute flux normalization $A=3.88  \times
10^6$~cm$^{-2}$s$^{-1}$MeV$^{-1}$ obtained by \citet{wsb99} from data
at $E=1$~TeV.  Alternative choices are available in the literature
\citep{gj67,gfr70,dww76,rm78}, but their agreement with this assumed
spectrum is good.  Our adopted spectrum also agrees well with the more
recent results quoted by \citet{w98}; discrepancies of around 50\% are
found below 100 MeV where solar wind modulation is important, but above
that the general agreement is within 10\%.  We do not include an
extended low energy tail that has been invoked in previous models, but
as we will show in section 4, the full treatment for evolution does not
require this. The values of ${\cal P}_{i,j\rightarrow \ell}$ for the
reactions considered, evaluated with the GCR spectrum and nuclear cross
section just described, are shown in Table~\ref{tab:Pijk}.  These
production coefficients contain all the necessary nucleosynthesis
information.

\section{Analytic Results Models for Light Element Evolution in a 
Closed Box Formalism} 
 
The simplest representation of Galactic evolution is the single zone
closed box scheme, also called the Simple Model \citep[SM, see {\em
e.g.}][]{t80}. It provides a useful, although limited, test bed for
model assumptions and here we use it as a guide to more complete
chemical evolution modeling that forms the next section.  The main
advantage of this closed box picture is that we can  derive analytical
expressions for the Li, Be and B abundances, evaluate orders of
magnitude of the most significant quantities, and gain an idea of their
trends.  It should be kept in mind, however, that the model is severely
restricted: there can be no net flow of mass through the zone, nor is
there coupling with any other parts of the Galaxy.

In the SM, the Galaxy is a spatially homogeneous, isolated {\em  
closed volume} of fixed total mass $M_0$. The gas mass $M_g$ decreases with 
time from star formation at a rate $\psi$, 
\beq 
\label{SMgas} 
\frac{dM_g}{dt}=-(1-R)\psi, 
\enq 
where $R$ is the average mass fraction, weighted by the initial mass 
function (IMF) $\phi(m)$, returned to the ISM by stars at the end of 
their evolution.  In the following, we employ the instantaneous  
recycling approximation (IRA), and assume that stars process interstellar gas  
and replenish it on a timescale shorter than the typical large scale  
dynamical timescales.   The rate of metal production by stars is: 
\beq 
\label{SMz} 
M_g\frac{dZ}{dt}=P_Z\psi, 
\enq 
where $Z$ is the metallicity and $P_Z$ is the mass fraction of heavy
elements newly produced and ejected by a stellar generation weighted by
the IMF.  As usual in such treatments, we make no distinction between
the stellar and interstellar abundances.  Here we adopt the numerical
values obtained by \citet{gpfp95} from the same stellar and
nucleosynthetic data adopted in the evolution model of \citet{fmpp92}
$R=0.21$ and $P_Z=7.9\times 10^{-3}$.  In keeping with the restrictions
of the IRA, massive stars are the principal nucleosynthetic agents and
since the IMF remains unaltered with time, both $R$ and $P_Z$ are held
constant.

In a closed system, since the total mass $M_0$ must remain fixed, $M_g$ and
$t$ are conjugate variables and eqs. (\ref{SMgas}) and (\ref{SMz}) yield
\beq 
Z=-\frac{P_Z}{1-R}\ln\left(\frac{M_g}{M_0}\right), 
\label{met}
\enq 
with $Z(0)=0$.  Notice that we can include the effects of global mass
loss through star-induced processes, i.e. winds, supernovae, etc., by
inserting here a term proportional to $\psi$.  The net effect is the
same as reducing $R$.  On the other hand, gas infall is not so easily
treated.  It actually violates the basic assumption of the closed box:
there is a source of matter from outside the system.  Thus $M_g$ and
$Z$ are {\em explicit} functions of time and must be found
numerically.

Light elements are mainly destroyed in stars during their evolution,
except for limited sources of specific isotopes (during the AGB stage
and possibly from some novae).  On the other side of the coin, they are
produced in the ISM at a rate
$r_\ell(t)\equiv\sum_{i,j}r_{i,j\rightarrow \ell}(t)$, therefore the
abundance (by number) $x_\ell$ of the element $\ell$ is determined by
the equation
\beq 
\frac{d}{dt}(x_\ell M_g)=-x_\ell\psi+r_\ell M_g,
\enq 
or, using equation (\ref{SMgas}), by the equation 
\beq 
\label{SMxk}
M_g\frac{dx_\ell}{dt}=-x_\ell R\psi+r_\ell M_g.  
\enq

To solve equation (\ref{SMxk}), it is necessary to evaluate the term
$r_\ell(t)$, using equation (\ref{rijk}).  Following convention we
assume that the normalized supernova rate is proportional to the SFR,
$\xi(t)\equiv \psi_{\rm SM}(t)/\psi_{\rm SN}(t_{\rm Gal})=
\psi(t)/\psi(t_{\rm Gal})$, making no distinction between supernovae
types I and II.  This is a good approximation because the stars that
end their life as SN II are high mass stars and then characterized by
short lifetimes with respect to the age of the Galaxy.  We only use
supernovae to scale the production rate for cosmic rays, not for the
overall metallicity of the gas (for which the two types are implicitly
combined in the IRA coefficients).  As we will describe in the next
section, the distinction between SN types {\it is} made explicitly in
the fully nonlinear chemical evolution models.  With our assumption
$x_i^{\rm GCR}(t)=x_i^{\rm ISM}(t)$, the production coefficients
$r_\ell$ for both direct and inverse processes scale linearly with the
gas metallicity $Z$, since the abundance of $p$ and $\alpha$ remains
{\it nearly} constant with time. For fusion $\alpha$--$\alpha$
reactions the production coefficient is practically independent on
$Z$.  Therefore, for {\em direct} $p$ and $\alpha$ induced reactions
and {\em inverse} CNO induced reactions, we obtain
\beq 
\label{colD} 
r_\ell(t)={\xi}(t)\frac{Z(t)}{Z(t_{\rm Gal})} r_\ell(t_{\rm
Gal})\equiv \frac{r_\ell(t_{\rm Gal})t_0}{M_0 Z(t_{\rm Gal})}\psi(t)Z(t),
\enq 
where we have defined a timescale $t_0=M_0/\psi(t_{\rm Gal})$.
For $^6$Li and $^7$Li we must also include {\em fusion} $\alpha$--$\alpha$
reactions,  
\beq \label{colF} 
r_\ell(t)=\xi(t)r_\ell(t_{\rm Gal})
\equiv\frac{r_\ell(t_{\rm Gal})t_0}{M_0}\psi(t).  
\enq 
Although the same analytic expression describes both direct and inverse
processes, keep in mind that under our assumptions the $Z$ dependence
in the first case represents the instantaneous interstellar abundances,
while in the second case it is the GCR composition.

With these definitions, it is straightforward to solve equation (\ref{SMxk})
together with equation (\ref{SMgas}) and (\ref{SMz}):  
\begin{eqnarray} 
\label{dirinv}
\lefteqn{x_\ell(Z)=x_{\ell,{\rm BB}}\exp{\left(-\frac{R}{P_Z}Z\right)}}
\nonumber \\ 
& & +\frac{r_\ell(t_{\rm Gal})t_0 P_Z}{(1-2R)^2Z(t_{\rm Gal})} 
\left[\exp{\left(-\frac{R}{P_Z}Z\right)}-\left(1+\frac{1-2R}{P_Z}Z\right) 
\exp{\left(-\frac{1-R}{P_Z}Z\right)} \right], 
\end{eqnarray} 
for direct and inverse processes, and 
\beq
\label{fus} 
x_\ell(Z)=x_{\ell,{\rm BB}}\exp{\left(-\frac{R}{P_Z}Z\right)}+
\frac{r_\ell(t_{\rm Gal})t_0}{1-2R}
\left[\exp{\left(-\frac{R}{P_Z}Z\right)}
-\exp{\left(-\frac{1-R}{P_Z}Z\right)}\right], 
\enq 
for fusion reactions.  Any contribution from Big Bang nucleosynthesis
is included as an initial abundance $x_{\ell,\rm{BB}}$.  Only for $^7$Li the
abundance depends on all of the processes we have discussed, including
a significant cosmological contribution.

Equations (\ref{dirinv}) and (\ref{fus}) show that the light elements
evolution is determined by the combination of stellar destruction,
which reduces the initial Big Bang abundance, and cosmic-ray production
that is proportional to $r_\ell(t_{\rm Gal})t_0$. Thus, according to
the SM, all the physics of Galactic evolution is parameterized by the
single quantity $t_0$ while the remaining quantities in equations
(\ref{dirinv}) and (\ref{fus}) depend only on stellar properties.

In the absence of infall, the star formation rate must exponentially
decrease with time since inert remnants inevitably result from stellar
evolution.  With this assumed rate, it is straightforward to solve 
eq.~(\ref{SMgas}) and (\ref{SMz}), obtaining
\beq
M_g(t)=M_0 e^{-t/\tau},
\enq
\beq
\psi(t)=\frac{M_0}{(1-R)\tau} e^{-t/\tau},
\enq
and
\beq 
Z(t)=\frac{P_Z}{(1-R)\tau}t.  
\label{zt}
\enq
From eq.~(\ref{zt}) we obtain $\tau\simeq 4.2$~Gyr by requiring that by
$t=t_\odot=8.5$~Gyr the metallicity has reached the solar value
$Z_\odot=0.02$. For $t_{\rm Gal}\simeq 13$~Gyr, eq.~(\ref{zt}) also gives
$Z(t_{\rm Gal})\simeq 1.5 Z_\odot$. 
Notice that \citet{s86} estimated $\tau\simeq 4$~Gyr. Then for 
$t_0$, we obtain
\beq 
t_0\equiv\frac{M_0}{\psi(t_{\rm Gal})}
=(1-R)\tau e^{t_{\rm Gal}/\tau}\simeq 72~{\rm Gyr}. 
\label{t0}
\enq 

The coefficients $r_\ell(t_{\rm Gal})$ are given by
eq.~(\ref{defrijk}).  The production coefficients ${\cal
P}_{i,j\rightarrow \ell}$ and the trapping fractions $S_{i,j\rightarrow
\ell}$ are given by Table~\ref{tab:trapping} and \ref{tab:Pijk},
respectively. For both GCR and ISM at $t=t_{\rm Gal}$ we have adopted a
solar composition, $x_{\rm H}=0.91$, $x_{\rm He}=0.089$, $x_{\rm
C}=3.30\times 10^{-4}$, $x_{\rm N}=1.02\times 10^{-4}$, and $x_{\rm
O}=7.73\times 10^{-4}$ \citep{gs98}.  The values of $r_\ell(t_{\rm
Gal})$ obtained in this way are listed in Table~\ref{tab:3}.  With
these, and the value of $t_0$ given by eq.~(\ref{t0}), we obtain the
curves shown in Figures 1--4.

For Li evolution, shown in Figures 1 and 2, the predicted abundance
remains very close to the primordial value, represented by the Spite
plateau, up to [Fe/H] $\simeq -1.4$. At higher metallicities the
predicted abundance increases about tenfold, although remaining
significantly below the upper envelope of the observational data and
the solar system meteoritic value.  We will see in Section 4 that a
more realistic model of Galactic evolution considerably reduces the
predicted Li abundance at high metallicity, thus highlighting the need
to include an additional (stellar) source of Li at later Galactic
epochs.  In Fig. 1, we see that the main contribution to the rise at
[Fe/H]$\geq -2$ comes from fusion reactions that depend linearly on the
relative fraction of GCR $\alpha$ particles.  If, as reported in {\it
e.g.} \citet{mde98} and \citet{wsb99}, these are depleted relative to
CNO by as much as a factor of 5 without a change in spectrum, then the
rise is delayed and the enrichment of Li reduced as displayed.  The SM
predictions for Be and B (Figures 3 and 4) produce reasonable fits to
the observed abundances near $Z\simeq Z_\odot$, but fail at low
metallicity.  This is simply seen from equation (\ref{dirinv}) which,
for $Z\ll Z_\odot$, gives the approximate behavior of the abundance of
Be and B at low metallicity:
\beq 
x_\ell(Z)\simeq 
\frac{1}{2}\frac{r_\ell(t_{\rm Gal})t_0}{P_Z Z(t_{\rm Gal})}Z^2,
\enq 
showing a {\em quadratic} dependence on $Z$, at variance with the 
nearly linear trend up to [Fe/H] $\simeq -1.3$ \citep[see][]{bal99} in the
observational data.  In contrast, Li ($=^6$Li+$^7$Li) is 
predicted to have a linear dependence with $Z$,
\beq 
x_{\rm Li} \simeq x_{\rm Li,BB} 
+\frac{r_{\rm Li}(t_{\rm Gal})t_0-Rx_{\rm Li,BB}}{P_Z} Z
\enq
or,
\beq
\left(\frac{\rm Li}{\rm H}\right)
\simeq \left(\frac{\rm Li}{\rm H}\right)_{\rm BB}
+k\left(\frac{Z}{Z_\odot}\right),
\label{appr}
\enq
where
\beq
k\simeq\frac{r_{\rm Li}(t_{\rm Gal})t_0 Z_\odot}{x_{\rm H} P_Z}, 
\label{trend}
\enq 
and $r_{\rm Li}(t_{\rm Gal})=r_6(t_{\rm Gal})+r_7(t_{\rm Gal}) \simeq
9.8\times 10^{-28}$~s$^{-1}$ for fusion reactions (see
Table~\ref{tab:3}).  In eq.~(\ref{trend}) we have neglected the small
effect of astration of primordial Li represented by the term
$-Rx_{7,{\rm BB}}$ in eq.~(\ref{appr}). The value of $k$ has been
determined observationally by \citet{ral00} in a relatively small
sample of carefully selected low metallicity stars \citep{rnb99}.  For
a parametric linear fit they find $k=(1.1\pm 0.7)\times 10^{-8}$.  With
$x_{\rm H}=0.91$ the SM with our parameters predicts $k\simeq 6.2\times
10^{-9}$.  This accordance indicates that the spallation scenario is
likely correct, not that the closed box model is necessarily
appropriate, because this expansion is correct only for near solar
abundances and, as we have seen, the SM predictions are not consistent
with each other.  We therefore pass to the more complex model and show
how the light element evolution can be better modeled, though at the
expense of a considerably more complex treatment for the Galaxy.

\section{Results for Light Element Evolution in a
Multizone-Multi Population Framework for Galactic Chemical Evolution}

We now consider the production of LiBeB nuclei in the framework of a
more complex model of Galactic evolution.  We recall briefly some
important features of the model adopted \citep[for a more detailed
discussion see][]{fmpp92,fmpd94}. This multiphase and multi-zone model
is able to follow the abundances of H, D, $^3$He, $^4$He, $^{12}$C,
$^{13}$C, $^{14}$N, $^{16}$O, $^{20}$Ne, $^{24}$Mg, $^{28}$Si,
$^{32}$S, $^{40}$Ca, and $^{56}$Fe and has recently been extended to
$s$- and $r$-process elements \citep{t99}.  The model follows the
interconnected evolution of three different regions in the Galaxy:
halo, thick disk, and  thin disk ({\em multizone} treatment). Relaxing
the IRA takes into account detailed stellar nucleosynthesis as well as
finite stellar lifetimes.  The multiphase approach considers the
different phases of the matter in the Galactic system: ({\em i}\/)
stellar populations; ({\em ii}\/) stellar remnants; ({\em iii}\/)
interstellar matter phases.  A fundamental feature of the model is
self-regulation through the internal processes that allow for the
transformation of matter from one phase to another.  These interactions
produce the time dependence of the total mass fraction in each phase
and the chemical abundances in the ISM and in stars.  The star
formation rate $\psi(t)$ is determined self-consistently by the
interaction among the different phases of matter. This is very
important because in the majority of the existing models, such as the
SM described in Sec.~3, the formal time dependence for the SFR is
assumed {\em a priori}.  The initial mass function adopted in this
model is based on the analysis of \citet{f91} of the fragmentation of
molecular clouds in the solar neighborhood, and is constant in space
and time.

We have included the nucleosynthesis processes responsible for the
evolution of LiBeB nuclei in this general Galactic model but we
emphasize that {\em no other modification has been made to the model
parameters or equations}.  The results we report here require the input
from our standard version that has been successfully used to analyze a
number of Galactic evolution problems \citep[see {\em e.g.}][]{sf95}.
The nonlinear evolution equations that determine the Li, Be and B
abundances $x_\ell$ (by number), including spallation and fusion process,
are:
\beq 
\label{gcrH}
\frac{dx_{\ell,H}}{dt}=r_{\ell,H}-x_{\ell,H}\frac{W_H}{g_H+c_H}, 
\enq 
\beq
\label{gcrT} \frac{dx_{\ell,T}}{dt}=r_{\ell,T}-x_{\ell,T}\frac{W_T}{g_T+c_T}
+f_H(x_{\ell,H}-x_{\ell,T})\frac{g_H}{g_T+c_T}, 
\enq 
\beq 
\label{gcrD}
\frac{dx_{\ell,D}}{dt}=r_{\ell,D}-x_{\ell,D}\frac{W_D}{g_D+c_D}+f_T(x_{\ell,T}-x_{\ell,D})
\frac{g_T}{g_D+c_D}, 
\enq 
where $\ell=^{6}$Li, $^7$Li, $^9$Be, $^{11}$B is the light element under
consideration.   Here $f_H$ and $f_T$ are, respectively, the infall rates for gas 
flowing from the halo into the thick 
disk, and from thick disk into the thin disk: $f_H=0.10$ and $f_T=0.0065$ 
are the choices made
for the solar neighborhood, that reproduce the mass ratio of the
visible mass in the three zones.  The quantities $g_L$ and $c_L$ are
the gas and cloud fractions in the halo, thick disk, thin disk, respectively, where
$L=H$, $T$, or $D$ denotes the Galactic zone.  The destruction terms in 
the three zones are represented by $W_L(t)=\sum_i W_{i,L}(t)$, where
\beq 
W_{i,L}(t)=\sum_j\int_{m_{\rm
min}}^{m_{\rm max}}\tilde{Q}_{ij}(m)
X_j[t-\tau_Z(m)]\psi_L[t-\tau_Z(m)]dm, 
\enq 
is the mass fraction injected in the ISM as element $i$ from stars
ending their evolution at the time $t$.  In order to compute the
nucleosynthesis production of a star of mass $m$, the astration matrix
$Q_{ij}(m)$ has been introduced, following \citet{TA}. The $\{ij\}$
element of this matrix represents the fraction of the star mass
initially in the form of chemical species $j$, transformed and ejected
as chemical species $i$, $Q_{ij}(m)=m_{ij}^{\rm exp}/m_j$. The quantity
$\tilde{Q}_{ij}(m)$ is obtained weighting the contribution of the
normal evolving stars, including stars exploding as SN II  and the
binary system ending as SN I.  As we discussed in the previous section,
for the closed box the two types have been combined for their
contributions to cosmic ray production but the full chemical evolution
model accounts for the distinctions between these two types of
supernovae.

In Figures \ref{lithium}--\ref{boron} we show the abundances predicted
by our model, obtained following the nucleosynthesis prescriptions
described in Section 2.  The different trends for the metallicity as
function of time contribute as well to the differences apparent as
shown in Figure~\ref{metallicity}.  In the multiphase model, $Z$ is not
a linear function of time.  In fact, the IRA is relaxed and the
consequences are particularly notable at low $Z$, when the finite
stellar lifetime is most important. After $t\simeq 2$~Gyr, (or $\log
Z\simeq -2.5$) the SM provides a good approximation of the numerical
model.  This is just the range, [Fe/H]$\ge -0.8$, where the abundances
of LiBeB elements predicted by the two schemes are most similar.  The
displacement is essentially in [Fe/H] for this reason.

Another important question for chemical evolution models is the
normalization in the predicted abundances of LiBeB elements. It is
common practice \citep[see for example][]{p93,fo99a,fo99b,vfc99} to
normalize the final abundances of GCR-only isotopes $^6$Li, $^9$Be,
$^{10}$B to the observed abundances at [Fe/H] = 0.  Consequently the
GCR yields of $^7$Li and $^{11}$B are scaled by a factor that is the
average of the three scaling factors of these isotopes.  We stress that
the results of our models, shown in Figures \ref{lithium}--\ref{boron},
are obtained without any {\em ad hoc} normalization.  The most evident
and relevant feature of the new curves is its improved agreement with
the observations: whereas the simple model yields a quadratic
dependence of [Be/H] and [B/H] with [Fe/H], the multiphase Galactic
evolution model more nearly approaches the observed approximately
linear trend for increasing values of metallicity.  Thus, the slope for
the abundance of Be and B vs. [Fe/H] is $\sim 2$ for the halo, $\sim
1.4$ for the thick disk, $\sim 0.7$ for the thin disk.  For comparison,
\citet{bal99} determined empirically for Be vs. [Fe/H] a best-fit slope
$\sim 1$--1.3 for $-3.0<$ [Fe/H] $<-1$, and $\sim 0.6$--0.7 for $-1.0<$
[Fe/H] $<0.1$.  It worth stressing this result because it is different
from the conclusions obtained previously by other authors \citep[see
for example][]{vfc99,fo99a}.  Even treating only the direct processes
\citep[i.e. the ``GCR standard'' as named in][]{vfc99}, it is possible
to reproduce quite well the observed trend of Be and B with [Fe/H].
From this trial we can conclude that in our scenario of Galactic
evolution the ``GCR standard'' does not lead to a quadratic dependence
of Be and B vs Fe, in contrast to, for example, \citet{vfc99} and
\citet{fo99a}.

The model reaches the solar photospheric B abundance
$\log$(B/H)$=-9.45$ \citep{gs98} at [Fe/H] = 0, whereas it exceeds the
solar Be abundance $\log$(Be/H)$=-10.85$ \citep{cmb75}.  This trend
also appears in other published results \citep[see {\em e.g.}][]{lal98} 
that overproduce Be relative to its meteoritic and photospheric values.
This discrepancy with the observed Be and B abundances at higher
metallicity may be partially due to stellar depletion.  The systematic
departures at more recent times of the numerical predictions from the
observations suggests this interpretation.   In Table 4 we compare the
model results at solar metallicity with the solar system data
\citep{hcr99}.  We remind the reader that the model results refer to
ISM abundances, whereas spectroscopic observations can only provide
stellar photospheric values.  A comparison of the theoretical curves
and the Be abundance at the time of formation to the Solar System
(meteoritic value) and in the ISM would represent a more appropriate
test of the model. Although the ISM value 4.6 Gyr ago is unknown, Howk
et al. (2000) provide a lower limit on the {\it contemporary}
interstellar $\log$B/H $> -9.60$.  If we assume that this is the
same difference as between the models and the stellar observations,
there are two possible explanations.  The more radical is that there
could be indications of a mixing mechanism -- presently unexplained --
for stars younger than about 1 to 2 Gyr.  This is based on the
metallicity at which the first systematic deviations appear in the Be
and B evolution models.  We re-iterate that he models have not been
normalized nor have any parameters used for the stellar population and
heavy metal evolution been altered to obtain agreements between the
observed and predicted light element abundances.  We favor a more
prosaic explanation:  that the Be and B abundances have been
systematically {\it underestimated} for the higher metallicity stars,
likely due the combined effects of NLTE and the ``missing opacity''
problem most recently outlined by \citet{bell01} and references
therein.  If so, assuming the appropriateness of our embedded
assumptions about the production rates, the models point to the
systematic correction that is needed and could also serve as a
constraint in the search for the responsible opacity sources {\it
without the need to invoke new production sources for the light
elements at late times in galactic history}.

Direct processes and the fusion reaction do not reproduce the observed
values for $^7$Li at high [Fe/H].  This is similar to the results of
the SM and also to those obtained in the literature.    It is well
known that the case of $^7$Li is more complex: lithium is not a
GCR-only element and stellar production cannot be ignored \citep[see
{\em e.g.}][]{tal01}.  For $^{11}$B/$^{10}$B, the models predict an
essentially constant ratio, 2.4, from spallation processes alone.  A
possible solution is the production of $^{11}$B by $\nu$-induced
spallation of $^{12}$C in SN II \citep{ww95} but this and other likely
contributing sources have not been included in the calculation.  From
Figures \ref{lithium}--\ref{boron}, we note a decrease of the slope of
the the halo and thick disk curves at low [Fe/H] in the cases of Be and
B.  This is in better agreement with the observational data in this
range than the results obtained including only direct processes.  The
contribution of inverse processes is more significant at low
metallicity, because during the first evolutionary phases of the Galaxy
the abundance of interstellar CNO target nuclei is very low.  For Li,
the contribution of inverse processes is marginal with respect to the
pre-Galactic Big Bang production.

\section{Discussion and Conclusions}

The SM is a closed box model with IRA that provides a useful test for
more complex models of Galactic evolution because its predicted
metallicity is the highest possible.  This is due to the approximate
treatment (IRA) of gas return, which maximizes metal production from
stars and the efficiency of its re-incorporation into stars, and to
the absence of gas dilution by infall.  Therefore, in the usual
abundance vs. [Fe/H] plots, SM predictions roughly determine the right
boundary of the region occupied by more complex models for which these
assumptions are relaxed; our multiphase model of Galactic evolution
relaxes the IRA approximation and allows infall of gas allowing us to
follow simultaneously the evolution of three Galactic zones (halo,
thick disk and thin disk).  This fact is evident from the comparison
of the results of the SM with those of our model shown in Figures 2, 3
and 4.   For the halo component, the two models give similar results.
This is because the halo behaves like a closed box, even though it
loses gas to form the thick disk.  Here the loss of gas is not what
matters but the fact that there is {\it no feedback} on the halo from
the other Galactic zones.  Thus, the LiBeB isotope productions in the
halo follow the prediction of the SM but are shifted to lowest values
of [Fe/H] because the IRA overestimates the actual metal production.
The slope of $\log({\rm Be/H})$ and $\log({\rm B/H})$ {\it vs.} [Fe/H]
for the halo is close to 2, the value for the SM.  Indeed, this
quadratic dependence of the abundance of these elements on metallicity
in the halo is a natural consequence of our halo picture. 

On the other hand, for the thick and thin disk, the IRA is not the only
difference with the SM and the couplings halo--thick disk, and thick
disk--thin disk play important roles which are, of course, ignored in
closed box models.  Thus, in addition to an overall rightward shift of
the abundances vs. [Fe/H] with respect to the SM the fact that the
thick disk receives chemically enriched gas from the halo, and the
thick disk from the thick disk, allows progressively shallower slopes
of the abundance curves vs. [Fe/H].  Our goal here is not much to
discriminate among different stellar populations in the Galaxy. Rather,
we deliberately loosely use the terms ``halo'', ``thick disk'' and ``thin
disk'' to describe coupled Galactic zones forming in sequential cascade,
and we emphasize the different behavior of such a complex system with
respect to a homogeneous one-zone model, no matter how sophisticated.

An additional complication in the interpretation of the results is the
customary (but unavoidable) way of plotting the abundance of chemical
species not as function of time, as they are calculated, but as
function of metallicity $Z$ or [Fe/H], to allow a comparison with
spectroscopic data. Whereas time and metallicity scale linearly in the
case of the SM and the use of either variable is equivalent, the same
is not true for our multiphase model. First, the relation of
metallicity with time is not linear.  Second, each Galactic zone
follows a different enrichment history. For example, during the first 3
Gyr of Galactic life (where [Fe/H] rises from minus infinity up to
about $-0.7$) the abundance of LiBeB elements is higher in the halo,
followed by the thick disk and the thin disk.  This simply reflects the
behavior of the star formation rates in the three zones (see Fig. 5 of
Pardi et al.  1995); for the same reason,  [Fe/H] follows the same
ordering (see also their Fig. 6), so that a given value of [Fe/H] is
reached first by the halo and later on by the thick disk and the thin
disk.  Since LiBeB abundances grow more rapidly than Fe with time,
however, their ordering in the abundance vs. [Fe/H] plot is reversed.
At a specified value of [Fe/H], the halo has the lowest abundance, and
the thin disk has the highest.  These facts should be kept in mind when
interpreting our results for LiBeB shown in Fig. 2--4.

In conclusion, we show that the multi-zone multi-population model we
used previously for other aspects of Galactic evolution produces quite
good agreement with the {\em linear} trend observed at low
metallicities as consequences of multi-zone coupling of a galactic
chemical evolution model and detailed accounting for the feedback
between phases without the instantaneous recycling approximation or
time dependent variations in the cosmic-ray spectrum or the initial
mass function.  No {\it special} mechanisms are needed to produce the
increase in abundances at late times in Galactic history ({\it e.g.}
novae producing excess Li, variations in the low energy part of the
cosmic ray spectrum, etc.).  Indeed, the {\it overproduction} of the
light elements in our calculations suggests that the abundances of
these elements derived from stellar photospheric observations may be
{\it underestimates}.  We argue that reported discrepancies between
theory and observations do not represent a nucleosynthetic problem, but
instead are the consequences of inaccurate treatments of Galactic
evolution.

\acknowledgments

We are grateful to Sam Austin for cross section data in advance of
publication and to Constantine Deliyannis, Brian Fields, Peter
Biermann, and the late Reuven Ramaty for discussions and
correspondence.  SNS was supported in part by NASA and thanks the
Osservatorio Astrofisico di Arcetri and the director, Franco Pacini,
for their kind invitations for several extended summer visits in 
2000--2001.  FF, DG, and SNS acknowledge support from the Italian Ministry
for the University and for Scientific and Technological Research
(MURST) through a COFIN-2000 grant.  We thank the (anonymous) referee
for helpful questions, especially those to which we openly respond in
the final section of this paper.

\clearpage

\begin{deluxetable}{llll}
\tablewidth{0pt}
\tablecaption{\sc Trapping fractions $S_{i,j\rightarrow \ell}$
\label{tab:trapping}}
\tablehead{
\colhead{$i$} & \colhead{$j$} & \colhead{$\ell$} 
& \colhead{$S_{i,j\rightarrow \ell}$} \\
}
\startdata
$p$      & C,N,O       & $^{6,7}$Li, $^{9}$Be, $^{10,11}$B & 1    \\
$\alpha$ & C,N,O       & $^{6,7}$Li, $^{9}$Be, $^{10,11}$B & 0.86 \\
$\alpha$ & $\alpha$    & $^{6,7}$Li                        & 0.46 \\
C,N,O    & $p, \alpha$ & $^{6,7}$Li, $^{9}$Be, $^{10,11}$B & 0.25 \\
\enddata
\end{deluxetable}

\begin{deluxetable}{llllll}
\tablewidth{0pt}
\tablecaption{\sc Production Coefficients ${\cal P}_{i,j\rightarrow \ell}$ 
(in 10$^{-25}$~s$^{-1}$)
\label{tab:Pijk}}
\tablehead{
\colhead{$i,j$ or $j,i$} & \colhead{$^6$Li} & \colhead{$^7$Li} 
& \colhead{$^9$Be} & \colhead{$^{10}$B}& \colhead{$^{11}$B} \\
}
\startdata
$p$,C       &  2.64  & 4.82  & 0.723  & 5.00  & 13.3 \\
$p$,N       &  4.21  & 2.52  & 1.08   & 2.94  & 5.93 \\
$p$,O       &  3.03  & 4.75  & 0.957  & 3.15  & 6.91 \\
$\alpha$,C  &  11.5  & 15.2  & 3.84   & 13.0  & 24.1 \\
$\alpha$,N  &  4.17  & 4.80  & 1.83   & 9.26  & 15.2 \\
$\alpha$,O  &  3.93  & 5.11  & 1.86   & 0.87  & 11.2 \\
$\alpha$,$\alpha$  & 0.74    & 1.94   & & & \\
\enddata
\end{deluxetable}

\begin{deluxetable}{llllll}
\tablewidth{0pt}
\tablecaption{\sc Production rates $r_{{i,j}\rightarrow \ell}(t_{\rm Gal})$
for the closed-box model (in $10^{-30}$~s$^{-1}$)
\label{tab:3}}
\tablehead{$i,j$ & $^6$Li & $^7$Li & $^9$Be & $^{10}$B & $^{11}$B} 
\startdata
direct processes &  &  &  &  &  \\
$p$,C          & 79.3   & 145    & 21.7   &  150      & 399     \\
$p$,N          & 39.0   & 23.3   & 10.0   &  27.2     & 55.0    \\
$p$,O          & 213    & 334    & 67.3   &  222      & 486     \\
total $p$,CNO  & 331    & 502    & 99.0   &  399      & 940     \\
\hline
$\alpha$,C         & 29.0 & 38.4 & 9.70 & 32.8   & 60.9 \\
$\alpha$,N         & 3.26 & 3.75 & 1.43 & 7.23   & 11.9 \\
$\alpha$,O         & 23.2 & 30.2 & 11.0 & 5.15   & 66.3 \\
total $\alpha$,CNO & 55.5 & 72.4 & 22.1 & 45.2   & 139  \\ 
$\alpha$,$\alpha$  & 270  & 707  &   &   &   \\
\hline
inverse processes &  &  &  &  &  \\
C,$p$         & 19.8  & 36.2  & 5.43  & 37.5   & 99.9  \\
N,$p$         & 9.77  & 5.85  & 2.51  & 6.82   & 13.8  \\
O,$p$         & 53.3  & 83.5  & 16.8  & 55.4   & 122   \\
total CNO,$p$ & 82.9  & 126   & 24.7  & 99.7   & 236   \\
\hline
C,$\alpha$         & 8.44   & 11.2  & 2.82  & 9.54   & 17.7  \\
N,$\alpha$         & 0.94   & 1.09  & 0.42  & 2.10   & 3.45  \\
O,$\alpha$         & 6.76   & 8.79  & 3.20  & 1.50   & 19.3  \\
total CNO,$\alpha$ & 16.1   & 21.1  & 6.44  & 13.1   & 40.5  \\
\enddata
\end{deluxetable}

\begin{deluxetable}{l l l l }
\tablewidth{0pt}
\tablecaption{\sc Solar System and ISM LiBeB Values Compared to Model
\label{tab:4}}
\tablehead{Value & ($^6$Li+ $^7$Li)/H & $^9$Be/H & ($^{10}$B + $^{11}$B)/H} 
\startdata
Chondrites ($^a$) & $5.1\times 10^{-10}$ & $2.1\times 10^{-11}$ 
& $3.3\times 10^{-10}$ \\
Orgueil ($^a$) & $2.0\times 10^{-9}$ & $2.6\times 10^{-11}$ & $7.6\times 10^{-10}$ \\
Solar Photosphere ($^b$) & $1.3\times 10^{-11}$ & $1.4\times 10^{-11}$ 
& $3.5\times 10^{-10}$ \\
ISM     & $>3.7\times 10^{-10}$ ($^c$) & \ldots  & $>2.5\times 10^{-10}$ ($^d$) \\
Model: [Fe/H]=0  & $1.1\times 10^{-9}$ & $6.0\times 10^{-11}$ 
& $5.3\times 10^{-10}$ \\
\enddata
\tablerefs{($^a$)~Hanon et al.~(1999), ($^b$)~Grevesse \& Sauval~(1998), 
($^c$)~Lemoine et al.~(1993), ($^d$)~Howk et al.~(2000)}
\end{deluxetable}

\clearpage

\begin{figure}
\plotone{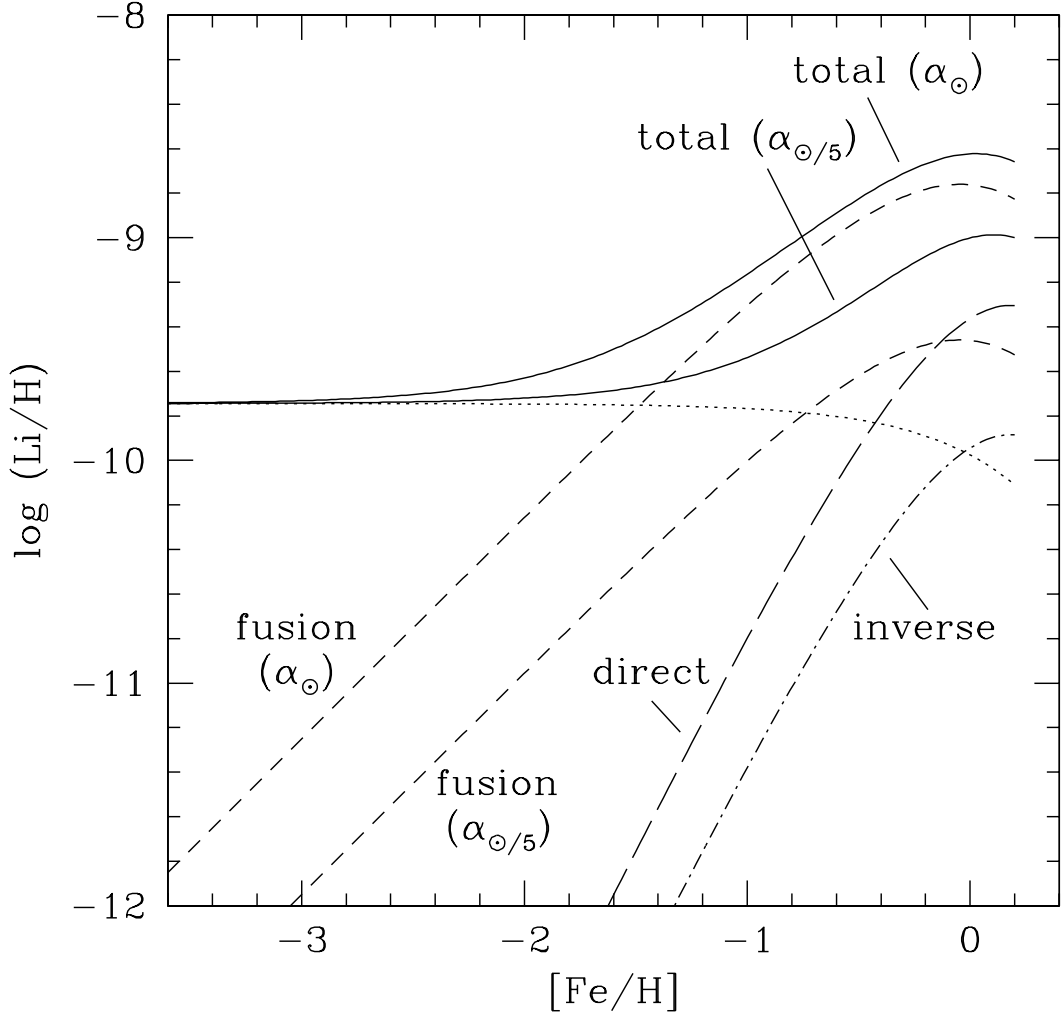}
\caption{Contributing physical processes to Li 
evolution ($=^6$Li+$^7$Li) according to the 
closed box model. 
{\em Dotted line}: Big Bang nucleosynthesis and stellar astration;
{\em short-dashed line}: fusion $\alpha$--$\alpha$ reactions;
{\em long-dashed line}: direct processes;
{\em dash-dotted line}: inverse processes.
The upper most 
curve shows the result obtained with a standard (solar) GCR composition 
as described in Sect. 3, the lower curve 
employed a reduced $\alpha$ contribution in the GCR flux by a factor of 5 (see 
text).}
\label{li-contr}
\end{figure}

\clearpage

\begin{figure}
\plotone{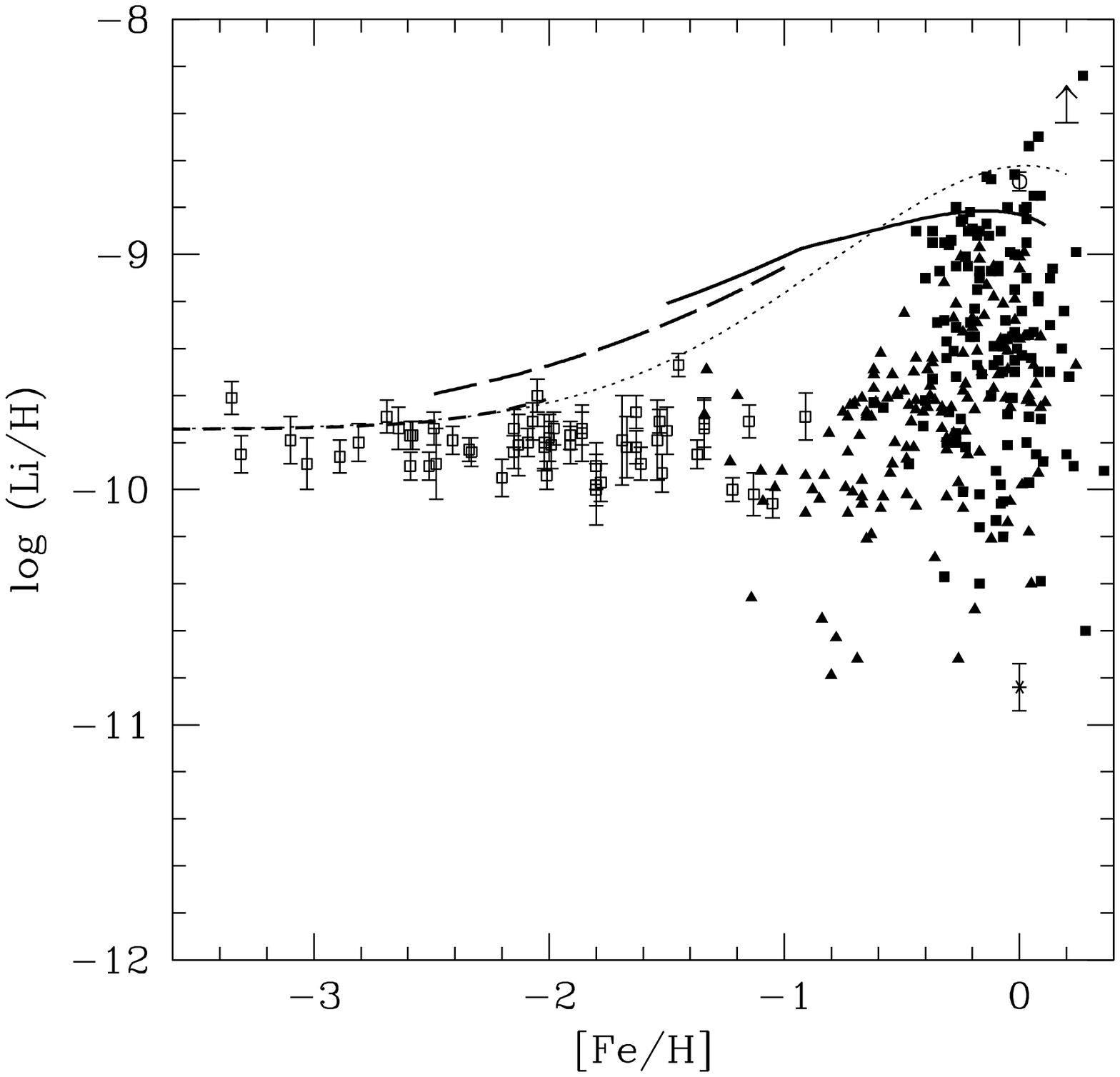}
\caption{Evolution of Li ($=^6$Li+$^7$Li) according to the closed box
model ({\em thin dotted line}) and our numerical model ({\em thick
short-dashed line}, halo; {\em thick long-dashed line}, thick disk;
{\em thick solid line}, disk) including the contribution of fusion,
direct and inverse reactions.  The meteoritic and the solar
photospheric Li abundances \citep[from][]{gs98} are indicated by an
{\em open circle} and an {\em asterisk}, respectively.  The lower limit
on the ISM abundance of Li determined by \citet{lal93} has been placed
at [Fe/H]$=0.2$.  Stellar photospheric data are from \citet{bm97} ({\em
open squares}), \citet{b90} ({\em filled squares}), \citet{f00} ({\em
filled circles}), and \citet{cal01} ({\em filled triangles}).}
\label{lithium}
\end{figure}

\clearpage

\begin{figure}
\plotone{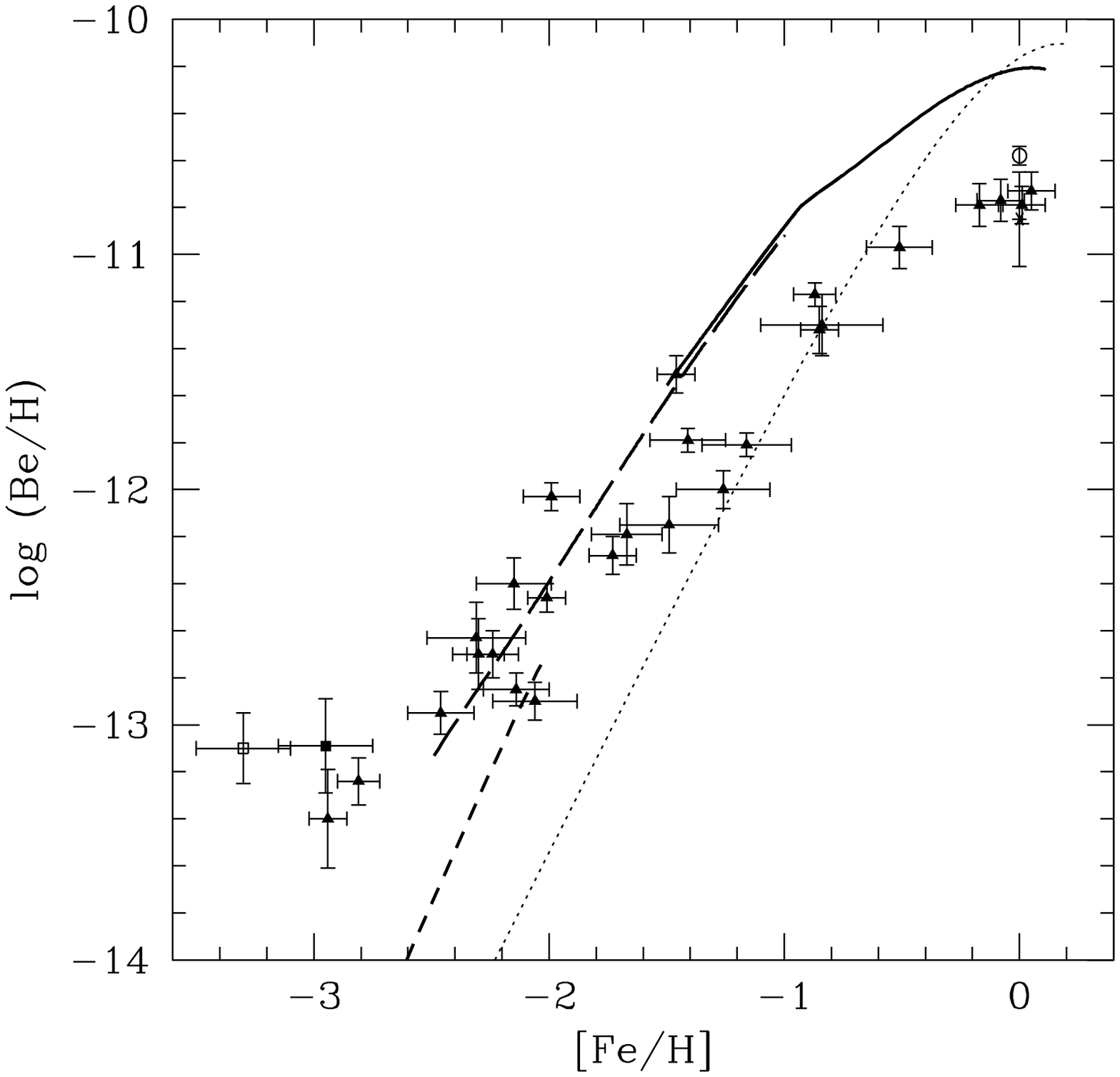}
\caption{Evolution of $^9$Be according to the closed box model ({\em
thin dotted line}) and our numerical model ({\em thick short-dashed
line}, halo; {\em thick long-dashed line}, thick disk; {\em thick solid
line}, disk) including the contribution of direct and inverse
reactions. The meteoritic \citep[from][]{gs98} and the solar
photospheric \citep[from][]{cmb75} $^9$Be abundances are indicated by
an {\em open circle} and an {\em asterisk}, respectively.  Stellar
photospheric data are from \citet{bal99} ({\em filled triangles}),
\citet{pal00a} (G64-12, {\em open square}), and \citet{pal00b}
(LP815-43, {\em filled square}).}
\label{beryllium}
\end{figure}

\clearpage

\begin{figure}
\plotone{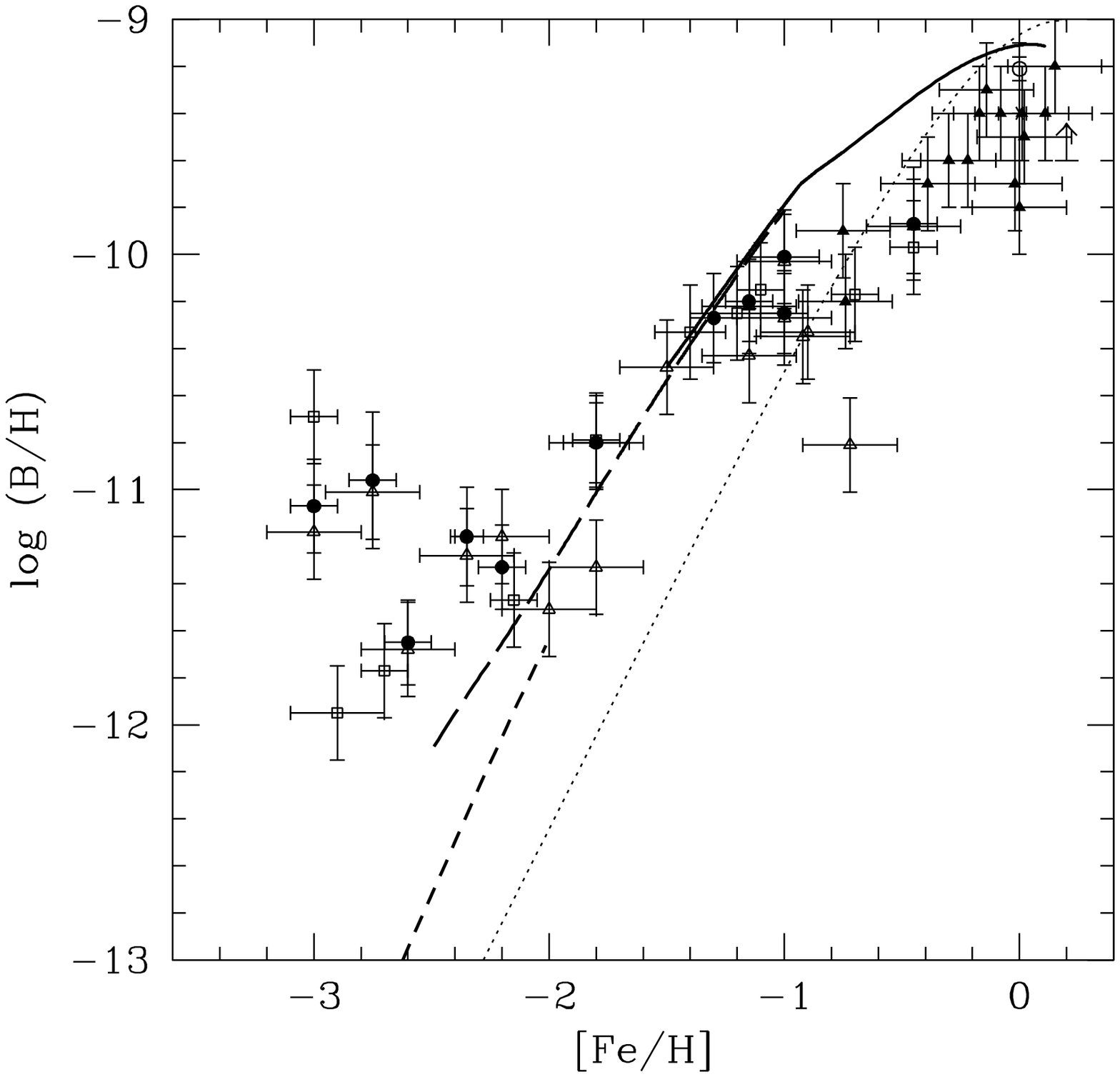}
\caption{Evolution of B ($=^{10}$B+$^{11}$B) according to the closed
box model ({\em thin dotted line}) and our numerical model ({\em thick
short-dashed line}, halo; {\em thick long-dashed line}, thick disk;
{\em thick solid line}, disk) including the contribution of direct and
inverse reactions. The meteoritic and the solar photospheric B
abundances \citep[from][]{gs98} are indicated by an {\em open circle}
and an {\em asterisk}, respectively.  The lower limit on the ISM
abundance of B determined by \citet{hss00} has been placed at
[Fe/H]$=0.2$.  Stellar photospheric data are from \citet{cal00} ({\em
filled squares}), \citet{pal99} ({\em open triangles}), \citet{dal97}
({\em filled circles}), and \citet{glal98} ({\em empty squares}).}
\label{boron}
\end{figure}

\clearpage

\begin{figure}
\plotone{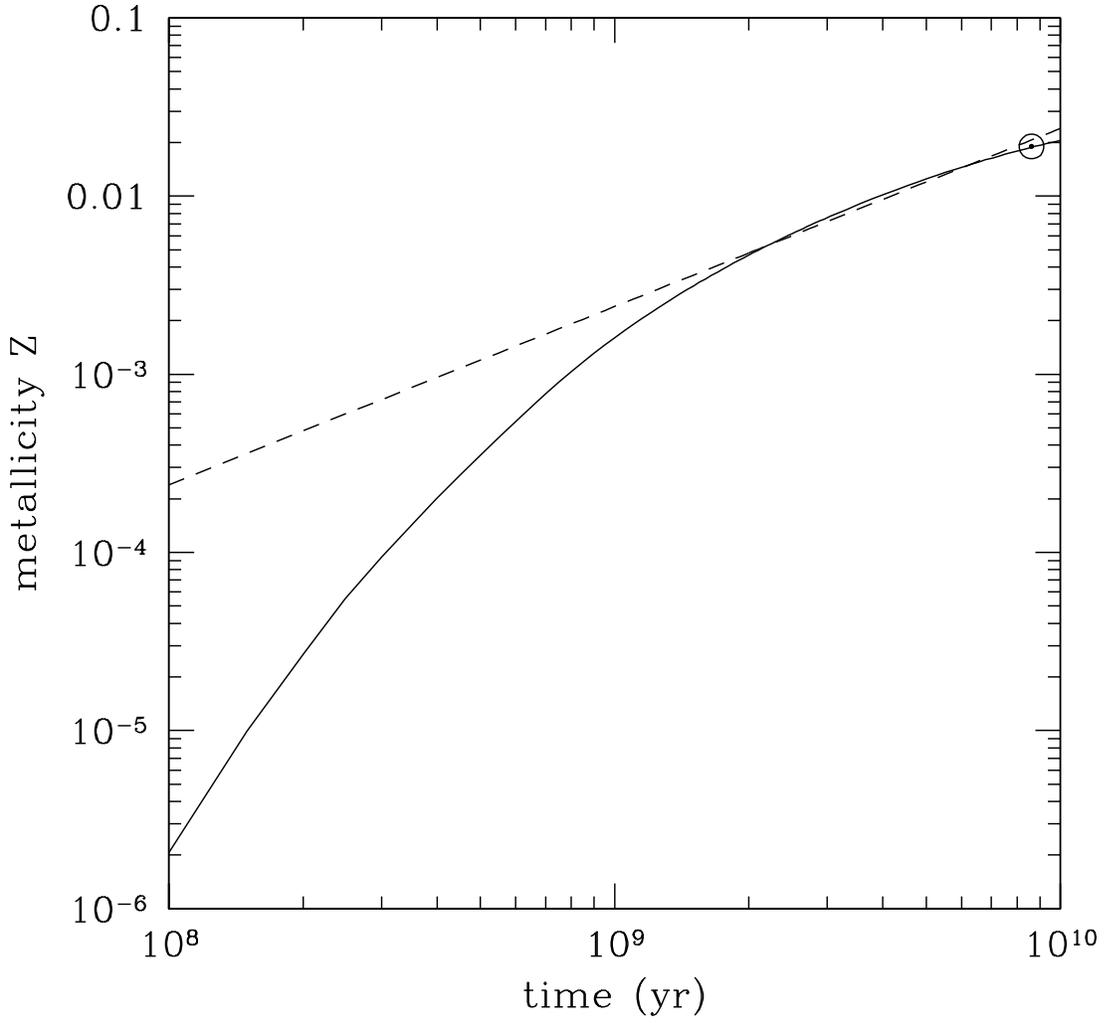}
\caption{Comparison of the metallicity $Z$ predicted by the SM ({\em
dashed line}) and by our model ({\em solid line}) as function of time.
The solar symbol indicates the solar metallicity at the epoch of
formation of the Sun, $t=8.5$~Gyr, assumed in our model.}
\label{metallicity}
\end{figure}

\end{document}